\documentclass[11pt]{article}
\usepackage{epsfig}
\begin{document}
\pagestyle{empty}

%

\newcommand{\bea}{\begin{eqnarray}}
\newcommand{\eea}{\end{eqnarray}}
\def\eqa{&=&}
\def\ccr{\nonumber\\}

\def\a{\alpha}
\def\b{\beta}
\def\m{\mu}
\def\n{\nu}
\def\r{\rho}
\def\s{\sigma}
\def\ep{\epsilon}

\def\sqr#1#2{{\vcenter{\vbox{\hrule height.#2pt
     \hbox{\vrule width.#2pt height#1pt \kern#1pt
           \vrule width.#2pt}
       \hrule height.#2pt}}}}
\def\square{\mathchoice\sqr66\sqr66\sqr{2.1}3\sqr{1.5}3}

\def\appendix{\par\clearpage
  \setcounter{section}{0}
  \setcounter{subsection}{0}
  \@addtoreset{equation}{section}
  \def\@sectname{Appendix~}
  \def\theequation{\thesection\arabic{equation}}
  \def\thesection{\Alph{section}}}

\def\thefigures#1{\par\clearpage\section*{Figures\@mkboth
  {FIGURES}{FIGURES}}\list
  {Fig.~\arabic{enumi}.}{\labelwidth\parindent\advance
\labelwidth -\labelsep
      \leftmargin\parindent\usecounter{enumi}}}
\def\figitem#1{\item\label{#1}}
\let\endthefigures=\endlist

\def\thetables#1{\par\clearpage\section*{Tables\@mkboth
  {TABLES}{TABLES}}\list
  {Table~\Roman{enumi}.}{\labelwidth-\labelsep
      \leftmargin0pt\usecounter{enumi}}}
\def\tableitem#1{\item\label{#1}}
\let\endthetables=\endlist

\def\@sect#1#2#3#4#5#6[#7]#8{\ifnum #2>\c@secnumdepth
     \def\@svsec{}\else
     \refstepcounter{#1}\edef\@svsec{\@sectname\csname the#1\endcsname
.\hskip 1em }\fi
     \@tempskipa #5\relax
      \ifdim \@tempskipa>\z@
        \begingroup #6\relax
          \@hangfrom{\hskip #3\relax\@svsec}{\interlinepenalty \@M #8\par}
        \endgroup
       \csname #1mark\endcsname{#7}\addcontentsline
         {toc}{#1}{\ifnum #2>\c@secnumdepth \else
                      \protect\numberline{\csname the#1\endcsname}\fi
                    #7}\else
        \def\@svse=chd{#6\hskip #3\@svsec #8\csname #1mark\endcsname
                      {#7}\addcontentsline
                           {toc}{#1}{\ifnum #2>\c@secnumdepth \else
                             \protect\numberline{\csname the#1\endcsname}\fi
                       #7}}\fi
     \@xsect{#5}}

\def\@sectname{}
%
%
\def\eg{\hbox{\it e.g.}}        \def\cf{\hbox{\it cf.}}
\def\etal{\hbox{\it et al.}}
\def\dash{\hbox{---}}
\def\bR{\mathop{\bf R}}
\def\bC{\mathop{\bf C}}
\def\eq#1{{eq. \ref{#1}}}
\def\eqs#1#2{{eqs. \ref{#1}--\ref{#2}}}
\def\lie{\hbox{\it \$}} 
\def\partder#1#2{{\partial #1\over\partial #2}}
\def\secder#1#2#3{{\partial^2 #1\over\partial #2 \partial #3}}
\def\abs#1{\left| #1\right|}
\def\ltap{\ \raisebox{-.4ex}{\rlap{$\sim$}} \raisebox{.4ex}{$<$}\ }
\def\gtap{\ \raisebox{-.4ex}{\rlap{$\sim$}} \raisebox{.4ex}{$>$}\ }
\def\contract{\makebox[1.2em][c]{
        \mbox{\rule{.6em}{.01truein}\rule{.01truein}{.6em}}}}
%
\def\com#1#2{
        \left[#1, #2\right]}
%
%
\def\bentarrow{\:\raisebox{1.3ex}{\rlap{$\vert$}}\!\rightarrow}
\def\longbent{\:\raisebox{3.5ex}{\rlap{$\vert$}}\raisebox{1.3ex}%
        {\rlap{$\vert$}}\!\rightarrow}
\def\onedk#1#2{
        \begin{equation}
        \begin{array}{l}
         #1 \\
         \bentarrow #2
        \end{array}
        \end{equation}
                }
\def\dk#1#2#3{
        \begin{equation}
        \begin{array}{r c l}
        #1 & \rightarrow & #2 \\
         & & \bentarrow #3
        \end{array}
        \end{equation}
                }
\def\dkp#1#2#3#4{
        \begin{equation}
        \begin{array}{r c l}
        #1 & \rightarrow & #2#3 \\
         & & \phantom{\; #2}\bentarrow #4
        \end{array}
        \end{equation}
                }
\def\bothdk#1#2#3#4#5{
        \begin{equation}
        \begin{array}{r c l}
        #1 & \rightarrow & #2#3 \\
         & & \:\raisebox{1.3ex}{\rlap{$\vert$}}\raisebox{-0.5ex}{$\vert$}%
        \phantom{#2}\!\bentarrow #4 \\
         & & \bentarrow #5
        \end{array}
        \end{equation}
                }
\newcommand{\nc}{\newcommand}
\nc{\spa}[3]{\left\langle#1\,#3\right\rangle}
\nc{\spb}[3]{\left[#1\,#3\right]}
\nc{\ksl}{\not{\hbox{\kern-2.3pt $k$}}}
\nc{\hf}{\textstyle{1\over2}}
\nc{\pol}{\varepsilon}
\nc{\tq}{{\tilde q}}
\nc{\esl}{\not{\hbox{\kern-2.3pt $\pol$}}}
\newcommand{\1}{{\'\i}}
\newcommand{\be}{\begin{equation}}
\newcommand{\ee}{\end{equation}\noindent}
\newcommand{\bear}{\begin{eqnarray}}
\newcommand{\ear}{\end{eqnarray}\noindent}
\newcommand{\benn}{\begin{enumerate}}
\newcommand{\enn}{\end{enumerate}}
\newcommand{\no}{\noindent}
\date{}
\renewcommand{\theequation}{\arabic{section}.\arabic{equation}}
\renewcommand{\arraystretch}{2.5}
\newcommand{\GeV}{\mbox{GeV}}
\newcommand{\cL}{\cal L}
\newcommand{\D}{\cal D}
\newcommand{\Dhalf}{{D\over 2}}
\newcommand{\Det}{{\rm Det}}
\newcommand{\PP}{\cal P}
\newcommand{\G}{{\cal G}}
\def\R{1\!\!{\rm R}}
\def\Eins{\mathord{1\hskip -1.5pt
\vrule width .5pt height 7.75pt depth -.2pt \hskip -1.2pt
\vrule width 2.5pt height .3pt depth -.05pt \hskip 1.5pt}}
\newcommand{\symb}{\mbox{symb}}
\renewcommand{\arraystretch}{2.5}
\newcommand{\slD}{\raise.15ex\hbox{$/$}\kern-.57em\hbox{$D$}}
\newcommand{\slpartial}{\raise.15ex\hbox{$/$}\kern-.57em\hbox{$\partial$}}
\newcommand{\slG}{{{\dot G}\!\!\!\! \raise.15ex\hbox {/}}}
\newcommand{\Gd}{{\dot G}}
\newcommand{\Gund}{{\underline{\dot G}}}
\newcommand{\Gdd}{{\ddot G}}
\def\GBd12{{\dot G}_{B12}}
\def\mneg{\!\!\!\!\!\!\!\!\!\!}
\def\Mneg{\!\!\!\!\!\!\!\!\!\!\!\!\!\!\!\!\!\!\!\!}
\def\non{\nonumber}
\def\beqn*{\begin{eqnarray*}}
\def\eqn*{\end{eqnarray*}}
\def\sy{\scriptscriptstyle}
\def\footstrut{\baselineskip 12pt}
\def\square{\kern1pt\vbox{\hrule height 1.2pt\hbox{\vrule width 1.2pt
   \hskip 3pt\vbox{\vskip 6pt}\hskip 3pt\vrule width 0.6pt}
   \hrule height 0.6pt}\kern1pt}
\def\np{n_{+}}
\def\nm{n_{-}}
\def\Np{N_{+}}
\def\Nm{N_{-}}
\def\exmn{\Bigl(\mu \leftrightarrow \nu \Bigr)}
\def\slash#1{#1\!\!\!\raise.15ex\hbox {/}}
\def\dint#1{\int\!\!\!\!\!\int\limits_{\!\!#1}}
\def\bra#1{\langle #1 |}
\def\ket#1{| #1 \rangle}
\def\vev#1{\langle #1 \rangle}
\def\rightvac{\mid 0\rangle}
\def\leftvac{\langle 0\mid}
\def\dps{\displaystyle}
\def\sy{\scriptscriptstyle}
\def\half{{1\over 2}}
\def\third{{1\over3}}
\def\fourth{{1\over4}}
\def\fifth{{1\over5}}
\def\sixth{{1\over6}}
\def\seventh{{1\over7}}
\def\eigth{{1\over8}}
\def\ninth{{1\over9}}
\def\tenth{{1\over10}}
\def\pa{\partial}
\def\ddtau{{d\over d\tau}}
\def\ge{\hbox{\textfont1=\tame $\gamma_1$}}
\def\gz{\hbox{\textfont1=\tame $\gamma_2$}}
\def\gd{\hbox{\textfont1=\tame $\gamma_3$}}
\def\go{\hbox{\textfont1=\tamt $\gamma_1$}}
\def\gt{\hbox{\textfont1=\tamt $\gamma_2$}}
\def\gth{\hbox{\textfont1=\tamt $\gamma_3$}}
\def\gf{\hbox{$\gamma_5\;$}}
\def\ie{\hbox{$\textstyle{\int_1}$}}
\def\iz{\hbox{$\textstyle{\int_2}$}}
\def\id{\hbox{$\textstyle{\int_3}$}}
\def\ldop{\hbox{$\lbrace\mskip -4.5mu\mid$}}
\def\rdop{\hbox{$\mid\mskip -4.3mu\rbrace$}}
\def\eps{\epsilon}
\def\epshalf{{\epsilon\over 2}}
\def\e{\mbox{e}}
\def\mn{{\mu\nu}}
\def\exmn{{(\mu\leftrightarrow\nu )}}
\def\ab{{\alpha\beta}}
\def\exab{{(\alpha\leftrightarrow\beta )}}
\def\g{\mbox{g}}
\def\kinb{{1\over 4}\dot x^2}
\def\kinf{{1\over 2}\psi\dot\psi}
\def\expk{{\rm exp}\biggl[\,\sum_{i<j=1}^4 G_{Bij}k_i\cdot k_j\biggr]}
\def\expp{{\rm exp}\biggl[\,\sum_{i<j=1}^4 G_{Bij}p_i\cdot p_j\biggr]}
\def\expshort{{\e}^{\half G_{Bij}k_i\cdot k_j}}
\def\expabb{{\e}^{(\cdot )}}
\def\epseps#1#2{\varepsilon_{#1}\cdot \varepsilon_{#2}}
\def\epsk#1#2{\varepsilon_{#1}\cdot k_{#2}}
\def\kk#1#2{k_{#1}\cdot k_{#2}}
\def\G#1#2{G_{B#1#2}}
\def\Gp#1#2{{\dot G_{B#1#2}}}
\def\GF#1#2{G_{F#1#2}}
\def\Dab{{(x_a-x_b)}}
\def\Dsq{{({(x_a-x_b)}^2)}}
\def\lag{( -\partial^2 + V)}
\def\PITD{{(4\pi T)}^{-{D\over 2}}}
\def\4piTD{{(4\pi T)}^{-{D\over 2}}}
\def\4piT4{{(4\pi T)}^{-2}}
\def\TintmD{{\dps\int_{0}^{\infty}}{dT\over T}\,e^{-m^2T}
    {(4\pi T)}^{-{D\over 2}}}
\def\Tintm4{{\dps\int_{0}^{\infty}}{dT\over T}\,e^{-m^2T}
    {(4\pi T)}^{-2}}
\def\Tintm{{\dps\int_{0}^{\infty}}{dT\over T}\,e^{-m^2T}}
\def\Tint{{\dps\int_{0}^{\infty}}{dT\over T}}
\def\pint{{\dps\int}{dp_i\over {(2\pi)}^d}}
\def\Dx{\dps\int{\cal D}x}
\def\Dy{\dps\int{\cal D}y}
\def\Dpsi{\dps\int{\cal D}\psi}
\def\Tr{{\rm Tr}\,}
\def\tr{{\rm tr}\,}
\def\sumij{\sum_{i<j}}
\def\freeexp{{\rm e}^{-\int_0^Td\tau {1\over 4}\dot x^2}}
\def\arraystretch{2.5}
\def\Ge{\mbox{GeV}}
\def\dA{\partial^2}
\def\DA{\sqsubset\!\!\!\!\sqsupset}
\def\FFdual{F\cdot\tilde F}
\def\mn{{\mu\nu}}
\def\rs{{\rho\sigma}}
\def\oplusotimes{{{\lower 15pt\hbox{$\scriptscriptstyle
\oplus$}}\atop{\otimes}}
}
\def\perppar{{{\lower 15pt\hbox{$\scriptscriptstyle \perp$}}\atop{\parallel}}}
\def\oopp{{{\lower 15pt\hbox{$\scriptscriptstyle
\oplus$}}\atop{\otimes}}\!{{\lo
wer 15pt\hbox{$\scriptscriptstyle \perp$}}\atop{\parallel}}}
%
%
\def\bbbr{{\rm I\!R}}
\def\bbbone{{\mathchoice {\rm 1\mskip-4mu l} {\rm 1\mskip-4mu l}
{\rm 1\mskip-4.5mu l} {\rm 1\mskip-5mu l}}}
\def\bbbz{{\mathchoice {\hbox{$\sf\textstyle Z\kern-0.4em Z$}}
{\hbox{$\sf\textstyle Z\kern-0.4em Z$}}
{\hbox{$\sf\scriptstyle Z\kern-0.3em Z$}}
{\hbox{$\sf\scriptscriptstyle Z\kern-0.2em Z$}}}}

\renewcommand{\thefootnote}{\protect\arabic{footnote}}
\hfill {\large AEI-2010-125}\\
\vspace{-10pt}

\hfill {\large ITP-UU-10/25}\\

\vspace{-10pt}
\hfill {\large SPIN-10/22}
\begin{center}
{\huge\bf Open string pair creation from worldsheet instantons}
\vskip1.3cm
{\large Christian Schubert$^{a,b}$ and Alessandro Torrielli$^{c}$
}
\\[1.5ex]

\begin{itemize}
\item [$^a$]
{\it
Max-Planck-Institut f\"ur Gravitationsphysik\\
 Albert-Einstein-Institut\\
M\"uhlenberg 1, D-14476 Potsdam, Germany
}
\item [$^b$]
{\it
Instituto de F\'{\i}sica y Matem\'aticas
\\
Universidad Michoacana de San Nicol\'as de Hidalgo\\
Edificio C-3, Apdo. Postal 2-82\\
C.P. 58040, Morelia, Michoac\'an, M\'exico\\
}
\item [$^c$]
{\it
Institute for Theoretical Physics and Spinoza Institute\\
Utrecht University\\
Leuvenlaan 4, 3584 CE Utrecht - The Netherlands\\
}

\end{itemize}
\vspace{1cm}
 {\large \bf Abstract}
\end{center}
\begin{quotation}
Worldline instantons provide a particularly elegant way to derive
Schwinger's well-known formula for the pair creation rate due to
a constant electric field in quantum electrodynamics.
In this note, we show how to extend this method to the
corresponding problem of open string pair creation.

\end{quotation}
\vfill\eject
\pagestyle{plain}
\setcounter{page}{1}
\setcounter{footnote}{0}

\vspace{20pt}
\section{Introduction: Schwinger's formula and its open string generalization}
\renewcommand{\theequation}{1.\arabic{equation}}
\setcounter{equation}{0}

It was realized already in the early days of quantum electrodynamics that
this theory implies the possibility of electron -- positron pair production
from the vacuum in a strong external electric field \cite{sauter,eulhei,schwinger}.
As shown by Schwinger \cite{schwinger}, the existence of this process
and the pair creation probability can be derived from the imaginary part
of the effective Lagrangian. For the case of a constant electric field of
magnitude
$E$, he obtained  the well-known formula (at the one-loop level)

\bear
{\rm Im} {\cal L}_{\rm spin}(E) &=&  \frac{(eE)^2}{8\pi^3}
\, \sum_{k=1}^\infty \frac{1}{k^2}
\,\exp\left[-\frac{\pi k m^2}{eE}\right]
\label{schwingerspin}
\ear
with $m$ the electron mass. Schwinger also gave the corresponding formula for
scalar quantum electrodynamics,

\bear
{\rm Im} {\cal L}_{\rm scal}(E) &=&  \frac{(eE)^2}{16\pi^3}
\, \sum_{k=1}^\infty (-1)^{k+1}\frac{1}{k^2}
\,\exp\left[-\frac{\pi k m^2}{eE}\right]
\label{schwingerscal}
\ear
The expressions (\ref{schwingerspin}),(\ref{schwingerscal}) are clearly
nonperturbative in the
field.

The corresponding problem for an open string moving in a constant
electromagnetic background field
was first considered by Burgess \cite{burgess}, who calculated the pair
creation rate in the weak field limit.
The full analogue of Schwinger's formulas was obtained, for both bosonic and
supersymmetric open strings,
by Bachas and Porrati \cite{bacpor}. For the bosonic open string, their result reads

\bear
\label{Sstringa}
&&{\rm Im}{\cal L}_{\rm string}(E) =\\
&&\, \, \, \, \, \frac{1}{4(2\pi)^{D-1}}\sum_{\rm states \,
S}\frac{\beta_1 + \beta_2}{\pi \epsilon}
\sum_{k=1}^{\infty}(-)^{k+1}
\Bigl(\frac{\abs{\epsilon}}{k}\Bigr)^{D/2}
\,{\rm exp}\Bigl(-\frac{\pi
k}{\abs{\epsilon}}(M_S^2+\epsilon^2)\Bigr)\nonumber
\ear
Here the first sum is over the physical states of the bosonic string, with
$M_S$ the mass
of the state. $D=26$ is the spacetime dimension. The parameters $\beta_{1,2}$
are defined
as

\bear
\beta_{1,2} &=& \pi q_{1,2} E
\label{defbeta}
\ear
where $q_{1,2}$ are the $U(1)$ charges at the string endpoints, and

\bear
\epsilon &=& \frac{1}{\pi} \Bigl({\rm arctanh}\beta_1+{\rm
arctanh}\beta_2\Bigr)
\label{defespilon}
\ear
The formula (\ref{Sstringa}) reproduces in the weak -- field limit Schwinger's
formula
for spin zero (\ref{schwingerscal}), as well as its generalizations to
arbitrary integer spin $J$.
For stronger fields it deviates from the field theory case, even qualitatively,
since, due to the rapid
growth of the density of string states the
total rate for pair production derived from (\ref{Sstringa})
diverges at a critical field strength \cite{bacpor}

\bear
E_{\rm cr} &=& \frac{1}{\pi \abs{{\rm max} \,q_i}}
\label{Ecr}
\ear
Heuristically, a field of this strength would break the string apart.
However, overcritical fields probably do make sense physically as a
mechanism for D-brane decay \cite{amss,dosato,Hahowa}.

Nowadays, there are many methods available to obtain Schwinger's formulas
(\ref{schwingerspin}), (\ref{schwingerscal}). Perhaps the most elegant one
is the worldline instanton method, which was invented by Affleck et al.
for the scalar QED case \cite{afalma} and generalized to spinor QED
in \cite{dunsch,dwgs}. It allows one to determine the
$k$th Schwinger exponent through the calculation of a single periodic
stationary trajectory. In the following, we will show how to extend this method
to the bosonic string case.

\section{The worldline instanton method}
\renewcommand{\theequation}{2.\arabic{equation}}
\setcounter{equation}{0}

For easy reference, let us begin with sketching the worldline
instanton calculation \cite{afalma}
of the spin zero Schwinger formula (\ref{schwingerscal}).

The (euclidean) one-loop effective action for scalar QED can be written in the
following
way \cite{feynman50}:

\bear
\Gamma_{\rm scal} [A] &=&
\int_0^{\infty}\frac{dT}{T}\, \e^{-m^2T}
\int_{x(T)=x(0)} {\mathcal D}x
\, \e^{-S[x(\tau)]} \non\\
S[x(\tau)] &=&
\int_0^Td\tau
\left(\frac{\dot x^2}{4} +i e A\cdot \dot x \right)
\label{PI}
\ear
Here $m$ is the mass of the scalar particle,
and the functional integral $\int {\mathcal D}x$ is over all closed spacetime
paths $x^\mu(\tau)$ which are periodic in the proper-time parameter $\tau$,
with period $T$.
Rescaling $\tau = Tu$, the effective action may be expressed as
\bear
&&\Gamma_{\rm scal} [A] =\\
&&\, \, \, \int_0^{\infty}\frac{dT}{T}\, \e^{-m^2T}
\int_{x(1)=x(0)} {\mathcal D}x
\, {\rm exp}\left[-\left(\frac{1}{4T}\int_0^1du \,
\dot x^2 +i e\int_0^1du \, A\cdot \dot x
\right)\right] \non
\label{PIscale}
\ear
where the functional integral $\int {\mathcal D}x$ is now over all closed
spacetime paths $x^\mu(u)$ with period $1$.
After this rescaling we can perform the proper-time integral using the
method of steepest descent.
The $T$ integral has a stationary point at
\bear
T_0 = \frac{1}{2m}\sqrt{\int_0^1 du \, \dot x^2}
\label{defT0}
\ear
leading to
\bear
{\rm Im}\, \Gamma_{\rm scal} =
{1\over m}\sqrt{\pi\over T_0}
\,{\rm Im} \int {\cal D}x \,
\e^{-\Bigl(m\sqrt{\int \dot x^2}
+ie\int_0^1 du A\cdot \dot x
\Bigr)}
\label{elTint}
\ear
Here we have implicitly used the large mass approximation
\bear
m\sqrt{\int_0^1 du\,\dot{x}^2}\gg 1 \quad .
\label{large}
\ear

The functional integral remaining in the effective action expression
(\ref{elTint}) may be approximated by a further, functional,
stationary phase approximation. The new, nonlocal, worldline ``action'',
\bear
S_{\rm eff} = m\sqrt{\int_0^1 du\, \dot x^2} + i e \int_0^1duA\cdot \dot x
\label{defS}
\ear
is stationary if the path $x_\alpha(u)$ satisfies
\bear
m{\ddot x_{\mu}\over \sqrt{\int_0^1 du\, \dot x^2}} &=& i e F_{\mn}\dot x_{\nu}
\label{statcond}
\ear
A periodic solution $x_\mu(u)$ to (\ref{statcond}) is called a ``worldline
instanton''.
Further, contracting (\ref{statcond}) with $\dot{x}_\mu$ shows that for such an
instanton
\bear
\dot{x}^2={\rm constant}\equiv a^2
\label{c2}
\ear
Generally, the existence of a worldline instanton for a background $A$
leads to an imaginary part in the effective action $\Gamma_{\rm scal}[A]$, and
the leading behavior is
\bear
{\rm Im}\Gamma_{\rm scal}[A]\sim e^{-S_0}\quad ,
\label{leading}
\ear
where $S_0$ is the worldline action (\ref{defS}) evaluated on the worldline
instanton.

For a constant electric background of magnitude $E$, pointing in the
$z$ direction, the Euclidean gauge field is
$A_3(x_4) = -iEx_4$. The instanton equation (\ref{statcond}) for this case can
be easily
solved, and the solutions are simply circles in the $z-t$ plane
of radius $\frac{m}{eE}$ \cite{afalma}:

\bear
x_k^3(u)=\frac{m}{eE}\,\cos(2 k \pi u)\quad , \quad
x_k^4(u)=\frac{m}{eE}\,\sin(2 k \pi u)
\label{circle}
\ear
(with $x_{1,2}$ kept constant).
The integer $k\in {\bf Z}^+$
counts the number of times the closed path is traversed, and
the instanton action (\ref{defS}) becomes

\bear
S_0 := S_{\rm eff}[x_k^{\mu}] &=&
2k\,\frac{m^2\pi}{eE} - k\,\frac{m^2\pi}{eE} = k\,\frac{m^2\pi}{eE}
\label{action-constant}
\ear
Thus in the large mass approximation (\ref{large})
the contribution of the instanton with winding number $k$  reproduces the
exponent of the $k$th term of Schwinger's formula (\ref{schwingerscal}).

\section{Generalization to the open string}
\renewcommand{\theequation}{3.\arabic{equation}}
\setcounter{equation}{0}
The one-loop effective action for an open string in an electromagnetic
background
field $A^{\mu}$ with constant field strength tensor $F_{\mu\nu}$ has, in
conformal gauge,
the following path integral representation
\cite{fratse_plb,acny,burgess,bacpor},

\bear
\Gamma[A] &=& \half \int_0^{\infty}{dT\over T}
(4\pi^2 T)^{-{D\over 2}}Z(T)
\int {\cal D}x \,\e^{-S_E[x,A]}
\label{L}
\ear
Here $T$ denotes the Teichm\"uller
parameter of the annulus, and
the path integral is over all the embeddings of the annulus at fixed $T$ into
$D=26$ dimensional flat space. The worldsheet action is

\bear
S_E \!\! &=& \!\!
\frac{1}{4\pi\alpha'}\int d\sigma d\tau \,\partial_a x^{\mu} \partial^a x_{\mu}
-i {q_1\over 2} \int d\tau\,
x^{\mu}\partial_{\tau}x^{\nu}F_{\mu\nu}\Bigr\vert_{
\sigma =0}
\non\\
&& -i {q_2\over 2} \int d\tau\,
x^{\mu}\partial_{\tau}x^{\nu}F_{\mu\nu}\Bigr\vert_{\sigma = \half}
\label{S}
\ear
Here $\alpha'$ is the Regge slope, which will be set equal to
$\half$ in the following. The worldsheet is parameterized as a rectangle
$\sigma \in \lbrack 0, \half\rbrack$ and $\tau \in \lbrack 0 , T\rbrack$ where
$\tau = T$ is identified with $\tau = 0$. We use euclidean conventions
where $\sigma^0 = -i\sigma^2 = -i\tau $, $x^0 = -ix^D$, and $A_D = - iA_0$.
$q_{1,2}$ are the
charges associated with the two boundaries. We will assume that $q_1 \ne q_2$,
which eliminates the M\"obius strip contribution to this amplitude.
$Z(T)$ is the partition function of oriented open-string states, which in terms
of the masses $M_S$ of these states is given by

\bear
Z(T) &=& \sum_{\rm oriented\,\, states} \e^{-\pi T M_S^2}
\label{defZ}
\ear
The equations of motion derived from (\ref{S}) are

\bear
(\partial_{\sigma}^2 + \partial_{\tau}^2)\, x^{\mu} &=& 0 \non\\
\partial_{\sigma}x^{\mu} &=& i\pi q_2 F_{\mn}\partial_{\tau}x^{\nu}\,\,
\qquad (\sigma = \half) \non\\
\partial_{\sigma}x^{\mu} &=& - i\pi q_1 F_{\mn}\partial_{\tau}x^{\nu}
\qquad (\sigma = 0) 
\label{eomS}
\ear
Let us now consider the constant electric field case,
$F_{D,D-1}=-F_{D-1,D} = iE$.
We use (\ref{defZ}) to rewrite

\bear
\Gamma[F] &=& \half
\sum_{\rm oriented\,\, states}
\int_0^{\infty}{dT\over T}
(4\pi^2 T)^{-{D\over 2}}
\e^{-\pi T M_S^2}
\int {\cal D}x \,\e^{-S_E[x,F]}
\non\\
\label{useZ}
\ear
We rescale $\tau = Tu$ and do the $T$ - integral by the method of steepest
descent.
The stationary point is

\bear
T_0 &=& \sqrt{\frac{I_u}{I_{\sigma} + 2\pi^2M_S^2}}
 \label{T0S}
\ear
where we have abbreviated

\bear
I_{\sigma} &:=& \int_0^1du\int_0^{\half}d\sigma \,\partial_{\sigma} x^{\mu}
\partial_{\sigma} x_{\mu} \nonumber\\
I_{u} &:=& \int_0^1du\int_0^{\half}d\sigma \,\partial_u x^{\mu} \partial_u
x_{\mu} 
\label{defI}
\ear
The new worldsheet action is

\bear
S_{\rm eff} &=& \frac{1}{\pi} \sqrt{I_u}\sqrt{I_{\sigma}+2\pi^2M_S^2}
-i {q_1\over 2} \int d\tau\,
x^{\mu}\partial_{\tau}x^{\nu}F_{\mu\nu}\Bigr\vert_{
\sigma =0}
\non\\
&& -i {q_2\over 2} \int d\tau\,
x^{\mu}\partial_{\tau}x^{\nu}F_{\mu\nu}\Bigr\vert_{\sigma = \half}
\label{newSstring}
\ear
It leads to the equations of motions (compare (\ref{eomS}))

\bear
\Bigl\lbrack I_u\partial_{\sigma}^2 + (I_{\sigma}+2\pi^2M_S^2)
\partial_{u}^2\Bigr\rbrack\, x^{\mu} &=& 0 \label{laplace}\\
T_0\partial_{\sigma}x^{\mu} &=& i\pi q_2 F_{\mn}\partial_{u}x^{\nu}\,\,
\qquad (\sigma = \half) \label{bchalf}\\
T_0\partial_{\sigma}x^{\mu} &=& - i\pi q_1 F_{\mn}\partial_{u}x^{\nu}
\qquad (\sigma = 0) \label{bczero}
\label{eomeff}
\ear
The $k$th worldsheet instanton solving these equations is obtained by the
following
ansatz:

\bear
x_k^{D-1} &=& N\cos(2\pi k u )\cosh(b-a\sigma) \non\\
x_k^D &=& N\sin(2\pi k u )\cosh(b-a\sigma)
\label{ansatz}
\ear
with the remaining coordinates constants.
We take equal signs for $k$ and $a$.
Then (\ref{laplace}) and (\ref{T0S}) imply that

\bear
T_0 &=& \frac{2\pi k}{a}
\label{condlaplace}
\ear
and eqs. (\ref{bchalf}), (\ref{bczero}) give

\bear
\sinh b &=& \pi q_1 E\cosh b \non\\
\sinh\Bigl(b-\frac{a}{2} \Bigr) &=& -\pi q_2 E\cosh\Bigl(b-\frac{a}{2}\Bigr) \
\label{condbc}
\ear
Eqs. (\ref{condbc}) determine the parameters $a,b$ as

\bear
b &=& {\rm arctanh} \beta_1 \label{b}\\
a &=& 2\Bigl({\rm arctanh}\beta_1+{\rm arctanh}\beta_2\Bigr)
\label{a}
\ear
Calculating $I_{u},I_{\sigma}$ we find

\bear
I_u &=& N^2 \frac{(2\pi k)^2}{2a}
\Bigl\lbrack \frac{a}{2} + \beta_1 \cosh^2b + \beta_2
\cosh^2(b-a/2)\Bigr\rbrack \nonumber\\
I_{\sigma} &=& \frac{a^2}{(2\pi k)^2} I_u - \half N^2a^2
\label{IuIsigma}
\ear
Finally, the combination of (\ref{IuIsigma}) with (\ref{T0S}) and
(\ref{condlaplace}) fixes the normalization
of the instanton:

\bear
N &=& \frac{2\pi M_S}{\abs{a}}
\label{N}
\ear
We can then evaluate the stationary action:

\bear
S_0 &=& S_{\rm eff} [x^{\mu}_k] = 2\pi^2M_S^2 \frac{k}{a}
\label{Sfin}
\ear
Noting that $a  = 2\pi\epsilon$
this correctly reproduces the exponent in (\ref{Sstringa}) in the large $M_S$
limit.

\section{Discussion}
\renewcommand{\theequation}{4.\arabic{equation}}
\setcounter{equation}{0}

Although our calculation does not provide new information on the string pair
creation problem,
we  consider it worth presenting nonetheless. This is because, in the QED case,
the worldline
instanton approach has turned out to offer a relatively easy route to obtain
pair creation
rates for certain classes of non-constant fields \cite{dunsch,dwgs,dunwan}.
Moreover,
the form of the critical trajectories may also provide new physical insights.
It would be interesting to extend this calculation to the prefactor
determinant, as well as to the superstring case. The method may possibly
also generalize to the problem of D-brane decay into open strings.

\bigskip

\no
{\bf Acknowledgements:} We thank O. Corradini, G.V. Dunne and Soo-Jong Rey for
helpful discussions.

\end{document}